\newcommand{\be}[0]{\begin{equation}}
\newcommand{\ee}[0]{\end{equation}}
\newcommand{\ba}[0]{\begin{eqnarray}}
\newcommand{\ea}[0]{\end{eqnarray}}

\documentclass{ws-ijmpa}
\usepackage{epsfig,multicol}
\newcommand\fverb{\setbox\pippobox=\hbox\bgroup\verb}
\newcommand\fverbdo{\egroup\medskip\noindent%
            \fbox{\unhbox\pippobox}\ }
\newcommand\fverbit{\egroup\item[\fbox{\unhbox\pippobox}]}
\newbox\pippobox
\begin{document}

\markboth{A. Mirjalili, M.Dehghani,  and M.M. Yazdanpanah} {Parton
densities with the quark  linear potential in the statistical
approach}

%
\catchline{}{}{}{}{}
%

\title{Parton densities with the quark  linear potential in
the statistical approach}

\author{Abolfazl Mirjalili}

\address{Physics Department, Yazd
University, 89195-741,Yazd, Iran\\
A.Mirjalili@Yazd.ac.ir}

\author{Majid Dehghani}

\address{Physics Department, Yazd University, 89195-741, Yazd, Iran\\
m.dehghani56@stu.yazd.ac.ir}

\author{Mohammad Mehdi Yazdanpanah}

\address{Faculty of Physics, Shahid Bahonar University of
Kerman, Kerman, Iran\\
Myazdan@uk.ac.ir}

\maketitle

\begin{history}
\received{Day Month Year}
\revised{Day Month Year}
\end{history}

\begin{abstract}
{The statistical approach is used to calculate the parton
distribution functions (PDFs) of the nucleon. At first it is
assumed that the partons are free particles and the light-front
kinematic variables are employed to extract the Bjorken
$x$-dependence of the PDFs. These PDFs are used to evaluate the
combinations of the sea quarks such as $\bar d-\bar u$. As our
first attempt to improve the result, we make the statistical
parameters to depend on $Q^2$, using different values of Gottfried
sum rule.  The related results are indicating better behavior by
accessing to the PDFs while they contain the $Q^2$ dependence
parameters. As a further task and in order to have more
improvement in the  calculations, a linear potential is considered
to describe the quark interactions. The solution of the related
Dirac equation  yields the Airy function and  is considered as a
wave function in spatial space. Using the fourier transformation
the wave functions are obtained in momentum space. Based on the
light-front kinematic variables and using a special method which
we call it ``k method", these functions can be written in terms of
the Bjorken $x$-variable. Following that   the statistical
features are accompanied with these functions. Considering an
effective approach which is used in this article, we do not need
to resort to any extra effects as were assumed in some articles to
get a proper results for PDFs. The obtained results for $\bar
d-\bar u$ and the $\frac{\bar d}{\bar u}$ ratio, using our
effective approach, are in good agreement with the available
experimental data and  some theoretical results.
}
\end{abstract}

\ccode{PACS numbers: 12.38.AW, 12.39.Ki, 12.40.Ee}

\section{Introduction}\label{sec-intro}
 One of the  important goals of particle physics
is the description of nucleons from the first principles, i.e.
determination of structure of nucleons (such as proton and
neutron) in terms of quark and gluon degrees of freedom
{\cite{Brodsky 94}}. To understand the reasons of breakdown of the
efforts in this direction, first we should know the existing
approaches to the problem and see where the difficulty lies. There
are two fundamental different approaches to describe the structure
of strongly interacting particles whose stable and natural states
are called hadrons:

Parton model: this model was first introduced by Feynman
{\cite{Feynman 69}and assumes that nucleon is composed of point
like constituents, named ``partons". Afterthen it was  clarified
that partons are in fact consists of the  quarks and gluons. This
model was later on named constituent quark model, and has many
phenomenological applications.

Quantum Chromodynamic (QCD): which is a SU(3) non-abelian quantum
field theory.After the discovery of asymptotic freedom, this model
became the accepted field theory to describe strong interactions.

Reconciling the above mentioned approaches has not been in a good
manner  successful up to now. The main reason for dilemma is that
at first QCD is based mainly on perturbative calculations. Because
of the presence of non-pertubative effects in hadronic physics,
methods to incorporate non-perturbative effects should be
invented. The front form of QCD {\cite{Gellman 64} appears to be
the best tool for this purpose. The front form is also one of the
basic tools for the statistical description of the  parton model
that is concerned in this paper, so it will be introduced briefly.
On this base, the deep inelastic lepton proton scattering can be
viewed as a sum of elastic lepton–parton scattering, in which the
incident lepton is scattered off a parton instantaneously and
incoherently. In the statistical approach the proton is assumed as
a equilibrium thermal system  made up of free partons (quarks,
anti-quarks and gluons) that they have energy and momentum due to
their temperature\cite{ma2009}.

The organization of the paper is as follows : Introducing the
statistical parton model is done in Sect. 2 and then some history
and applications of front form are given there . Basic tools to
enable us to employ the statistical approach are introduced in
Sect.3. We present our results in Sect.4, where with the use of
different values of Gottfried sum rule ($S_G$) at some energy
scales, we achieve to the energy dependence of parton
distributions in the employed statistical model. In continuation,
relativistic quantum mechanical view is accompanied with the
statistical approach. In this case quantum states of partons as
fermions are obtained in Sect.5, using the solution of the Dirac
equation under a linear potential which is proper to describe  the
typical interaction of the particles inside the nucleon. Following
this strategy, the results for partons as a function of Bjorken
$x$-variable can be calculated analytically which is done in
Sect.6. We then resort to an effective approach to employ the
quantum effect of the statistical approach in Sect.7 without
resorting to use some extra effects which have been assumed  in
Ref.[\refcite{tomio}] . The results especially for $\bar d-\bar u$
and the $\frac{\bar d}{\bar u}$ ratio are in good agreement with
the available experimental data and the   theoretical results in
Ref.[\refcite{tomio}]. We finally give our conclusions in Sect.8.
\section{Parton model and the statistical approach}
In the parton model approach, the nucleon structure functions are
described in terms of parton distribution functions . As stated
above, because of the non-perturbative effects it is not possible
to calculate the PDF's completely from the perturbative part of
the field theory, relating to the  strong interactions. This is a
sign that we are inevitable to  investigate other models which can
be described properly both the perturbative and non-perturbative
parts of QCD. Since quarks and gluons are confined in a volume of
nucleon size, it is expected that statistical properties are
important in determining parton distribution functions
{\cite{ma2009,Cleymans 88,Mac 89} . So the statistical parton
model is introduced to incorporate the effects which are related
to the properties of  partons. In the primarily  statistical
parton model it is assumed that the nucleon is a gas of
non-interacting valence quarks, sea quarks, anti-quarks and gluons
with a thermal equilibrium.

It is explained  that the proper language for high energy region
is light-front dynamics. Light front dynamics is related to
infinite momentum frame (IMF). It is not a straight forward job to
transit from ordinary instant-form approximation to front form
dynamics. Instant-form approximation is suitable for non
relativistic dynamics and assumes that the system is prepared in
an instant of time in its rest frame, and evolves to later time.
As is apparent it is not suitable for relativistic dynamics. The
front form is the suitable language and because of its importance
for statistical parton model, we insist on using it. To describe
the front form, let us have a review on the historical progress of
the parton model.

The parton model was first initiated by Feynman {\cite{Feynman
69}}. Then the scaling behavior of partons in  deep inelastic
scattering was indicated by  Bjorken {\cite{Bjor 69}} in the limit
of  infinite momentum frame. Dirac {\cite{Dirac 49}} was the first
who introduced the concept of light-front form for relativistic
dynamics, which he called ``front form'' . Another important
application of light-front dynamics is in the  deep inelastic
scattering processes. Later on Weinberg developed a new
formulation for the perturbative field theory in the IMF
{\cite{Wein 66}} and  it was realized that IMF and light front
dynamics are equivalent and scaling behavior can be understood
more easily in the light-front form {\cite{Zhang 94}} .

In an another approach to deal with the parton model, impulse
approximation is used in deep inelastic lepton-nucleon scattering
in perturbative  field theory in  the IMF {\cite{Ma 90}}. Impulse
approximation can be used when the time of interaction between the
projectile and the constituent of the target is much smaller than
the life time of the target as seen by projectile {\cite{Ma 90}} .
In hadronic physics the light transit time is comparable to the
time scale governing internal motions so it might seem that only a
fuzzy picture of instantaneous state of the hadron can be obtained
{\cite{Kogut 73}}. The solution is to study the hadron in a
reference frame with the speed near that of light.  In this case
the time dilation effect slows the internal motions so that
impulse approximation can be used.

\section{Basic concepts of the statistical model}

By the assumption that nucleon is a system of partons in thermal
equilibrium, the mean number of partons (in nucleon rest frame )
is given by \cite{Ma 08,sym}:
\begin{equation}\label{mean no}
\bar{N}_f= \int f(k^0) d^3 k\;,
\end{equation}
where $f(k^0)$ is a distribution function (Fermi-Dirac or
Bose-Einestein distribution):

\begin{equation}\label{dis fun}
f(k^0)=\frac{g_f V}{(2 \pi)^{3}}
\frac{1}{e^{\frac{k^0-\mu_f}{T}}\pm 1}\;.
\end{equation}

In this equation the plus sign is referring to  fermions (quark,
anti-quark) and the minus sign is for the bosons (gluon). The
$g_f$ is the degree of color-spin degeneracy (6 for quarks and
anti-quarks, 16 for gluon). Chemical potential is representing by
$\mu_f$. As is known from thermodynamic the sign for anti-particle
is opposite to  that of particle ($ \mu_{\bar{q}}=-\mu_q$), and
for massless bosons the chemical potential is zero ( $\mu_g=0 $).

The four vector of energy-momentum in Eq.(\ref{mean no}) is
defined by:
\begin{equation}\label{dis fun}
 \vec{k}=(k^1,k^2,k^3),\; k^0=\sqrt{\vec{k}^2+{m_f}^2}\;,
 \end{equation}
where $k^0$ is the  energy, $\vec{k}$ is 3-momentum and ${m_f}$ is
mass of parton. By imposing the  on-shell condition on
Eq.(\ref{mean no}), we will get:
\begin{equation}\label{mean onshell}
\bar{N}_f= \int f(k^0) \delta (k^0-
\sqrt{(k^3)^2-{k_{\perp}}^2+{m_f}^2} \;dk^0dk^3 d^2k_\perp\;,
\end{equation}

To transform them to light-front kinematic variables in the
nucleon rest frame, the following transformations are used:

\begin{equation}\label{light var}
k^+=k^0+k^3 ,\; \;k_\perp=(k^1,k^2),\; \;\; k^-=k^0-k^3 ,\; \;\
k^+=P^+ x=Mx\;,
\end{equation}
where $x$ is the light-front momentum fraction of nucleon carried
by parton and $M$ is the mass of nucleon. Using the light-front
variables the delta function and integration measure changes to:
\begin{eqnarray}\label{light chan1}
&&\delta(k^0- \sqrt{(k^3)^2-{k_{\perp}^{2}}+{m_f}^2})= 2 k^0
\theta(k^0)\delta(k^2 -{m_f}^2)\nonumber\\
&=&[1+\frac{k_\perp^2+m_f^2}{Mx^2}]\;\theta(x)\;\delta(k^-
-\frac{k_{\perp}^2+m_f^2}{Mx})\;,
\end{eqnarray}

\begin{equation}\label{light chan2}
dk^0 dk^3 d^2k_{\perp}=\frac{1}{2} M dk^- dx d^2k_{\perp}\;.
 \end{equation}
Therefore Eq.(\ref{mean onshell}) will appear as:
\begin{equation}\label{mean light}
\bar{N}_f= \int f(x,k_\perp)  dx d^2k_{\perp},
\end{equation}
where $f(x,k_\perp)$ is resulted by substituting the
Eq.(\ref{light chan2}) in  Eq.(\ref{mean onshell}) and integrating
it with respect to $k^-$:

\begin{equation}\label{finteg}
f(x,k_\perp) = \frac{g_f MV}{2(2\pi)^3} \frac{1}{exp(\frac{1/2(
Mx+ \frac{K_\perp ^2 +{m_f}^2}{Mx} )-\mu_f}{T})\pm 1}
[1+\frac{K_\perp ^2 +{m_f}^2}{(Mx)^2} ]\theta(x),
\end{equation}

Integrating  Eq.(\ref{finteg})  with respect to $k_\perp$ (with
the assumption of being isotropic in transverse plane), will yield
us:

\begin{eqnarray}\label{disfun}
f(x) &=& \frac{g_f MTV}{8\pi^2} \{(Mx+\frac{m_f ^2}{Mx}) Ln[1 \pm
Exp(-\frac{1/2(Mx+\frac{m_f ^2}{Mx})-\mu_f}{T})]\nonumber\\
&&- 2 T Li_2 (\mp Exp(-\frac{1/2(Mx+\frac{m_f
^2}{Mx})-\mu_f}{T}))\},
\end{eqnarray}
where upper sign is denoting to fermions, negative sign to bosons
and $Li_2$ is the polylogarithm function. Eq.(\ref{disfun}) is
representing  the parton distribution function whose free
parameters will be determined, using the related constraints. The
statistical parameters of the proton, will be obtained if the
following sun rules are fulfilled:
\begin{equation}\label{sumrule1}
u_v=\int_{0}^{1} [u(x)-\bar{u}(x)] dx=2,
\end{equation}
\begin{equation}\label{sumrule2}
d_v=\int_{0}^{1} [d(x)-\bar{d}(x)] dx=1,
\end{equation}
\begin{equation}\label{sumrule3}
\int_{0}^{1} x[u(x)+\bar{u}(x)+d(x)+\bar{d}(x)+g(x)] dx=1\;.
\end{equation}
The free parameters of PDF  in Eq.(\ref{disfun}) are T, V, $\mu_u$
and $\mu_d$. As is proposed in {\cite{ma2009}}, the value of T is
known and the other parameters will be determined, using
Eqs.(\ref{sumrule1},\ref{sumrule2},\ref{sumrule3}). To obtain the
four unknown  parameters in Eq.(\ref{disfun}) we need to an extra
constrain in addition  to the existed  sum rules  for partons.
This extra constrain is related to   a reasonable value for $S_G$
in correspond to the available experimental data. Therefore there
are four parameters which need be obtained while the  four existed
equations  should be solved simultaneously. The $S_G$ is given by:
\begin{equation}\label{gottfried}
S_G=\int_{0}^{1} \frac{F_2^p-F_2^n}{x} dx= \frac{1}{3}+
\frac{2}{3} \int_{0}^{1} [\bar{u}(x)-\bar{d}(x)] dx,
\end{equation}
The experimental result for Gottfried sum rule is $S_G=0.235 \pm
0.026 $  {\cite{EMC}}. The authors in \cite{ma2009} have found the
Gottfried sum rule at value T=47 $MeV$, equals to $S_G=0.236$
which agrees well with the experimental data. The other three
parameter values are: $V=1.2\times 10^{-5}$ $MeV^{-3}$, $\mu_u=64$
$MeV$ , $\mu_d=36$ $MeV$.

\begin{figure}
\begin{tabular}{cc}
\centering
\begin{minipage}{6.2cm}
{\hspace{0.5 cm}\psfig{file=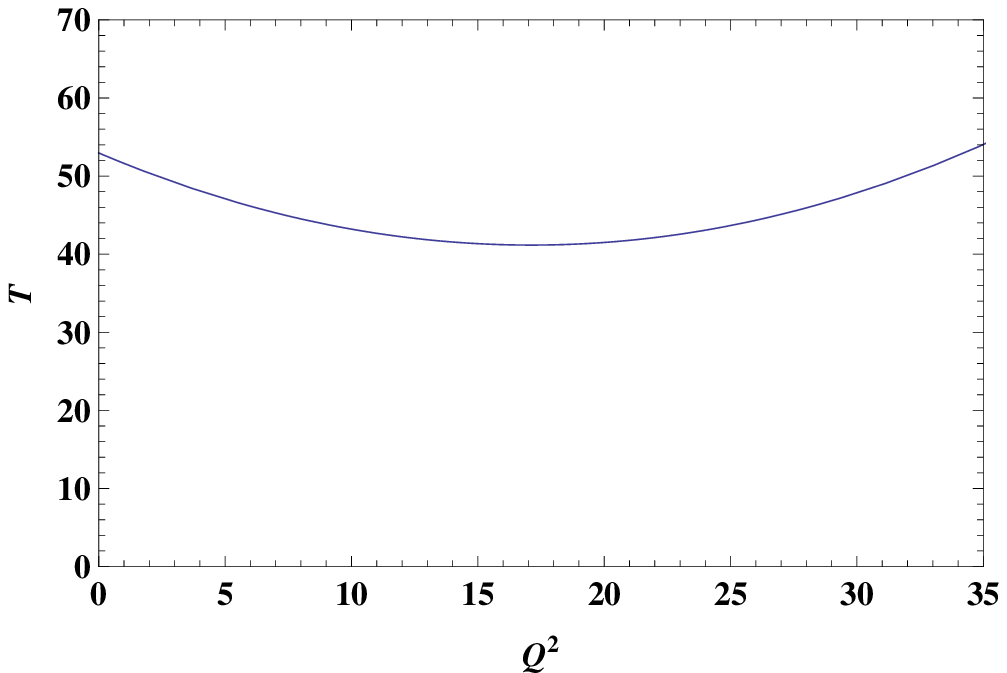,width=6cm}}
\caption{\hspace{-0.3 cm}Temperature parameter, T,verses $Q^2$
.\label{f1}}
\end{minipage}
&\;\;\;\;
\begin{minipage}{6.2cm}
{\psfig{file=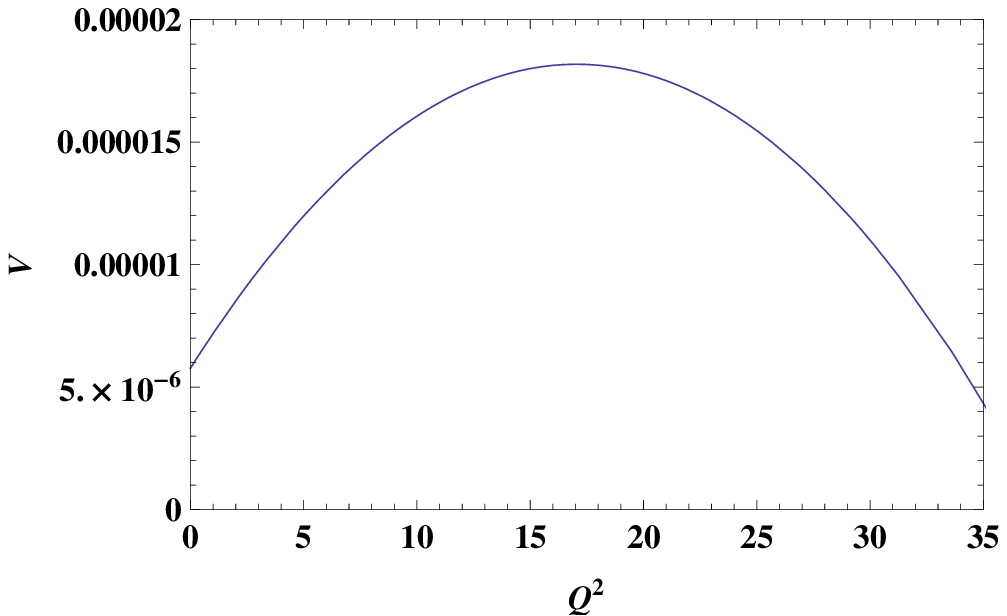,width=7cm}}  \caption{\hspace{-0.1
cm}Volume parameter, V,verses $Q^2$ .\label{f2}}
\end{minipage}
\end{tabular}
\end{figure}
\section{Energy dependence of parameters}

With the $x$-dependent distribution function which we access it
according to Eq.(\ref{disfun}) and solving the set of
Eqs.(\ref{sumrule1}-\ref{gottfried}) simultaneously, the four
parameters of statistical parton model, T, V, $\mu_u$ and $\mu_d$
can be computed at desired energy scales which can be related to
the $S_G$ values at some available energy scales, quoted in
[\refcite{Sg,AF}]. In Ref.[\refcite{AF}] the required plot of
$S_G$ against $Q^2$ has been plotted.

\begin{figure}
\begin{tabular}{cc}
\centering
\begin{minipage}{6.2 cm}
{\psfig{file=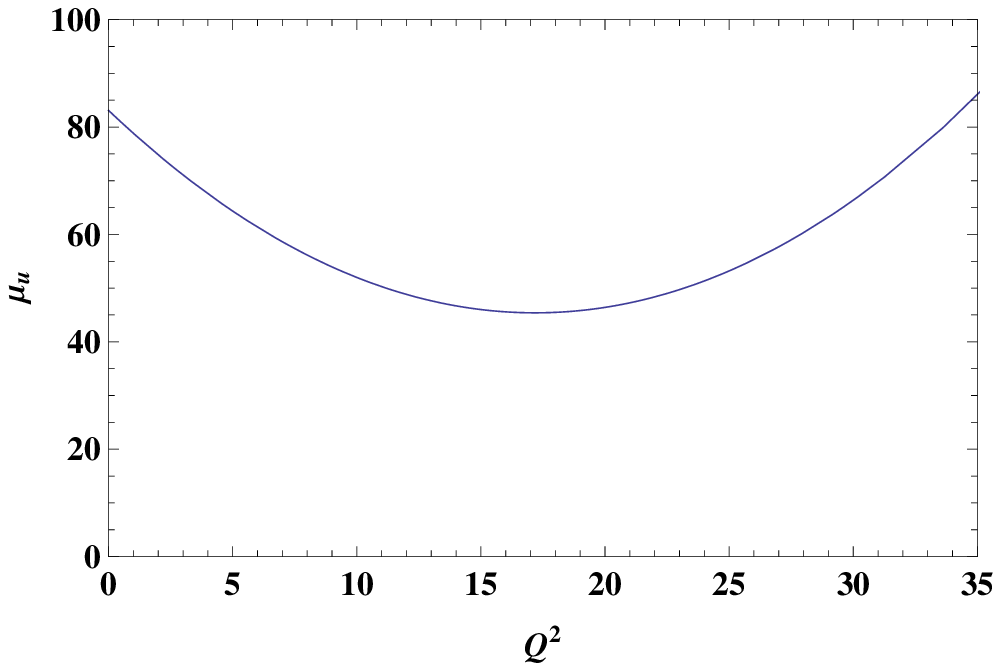,width=6 cm}} \caption{Chemical
potential for up quark, $\mu_u$, verses $Q^2$.\label{f3}}
\end{minipage}
&\;\;\;\;\;\;\;\;\;\;
\begin{minipage}{6.2 cm}
{\psfig{file=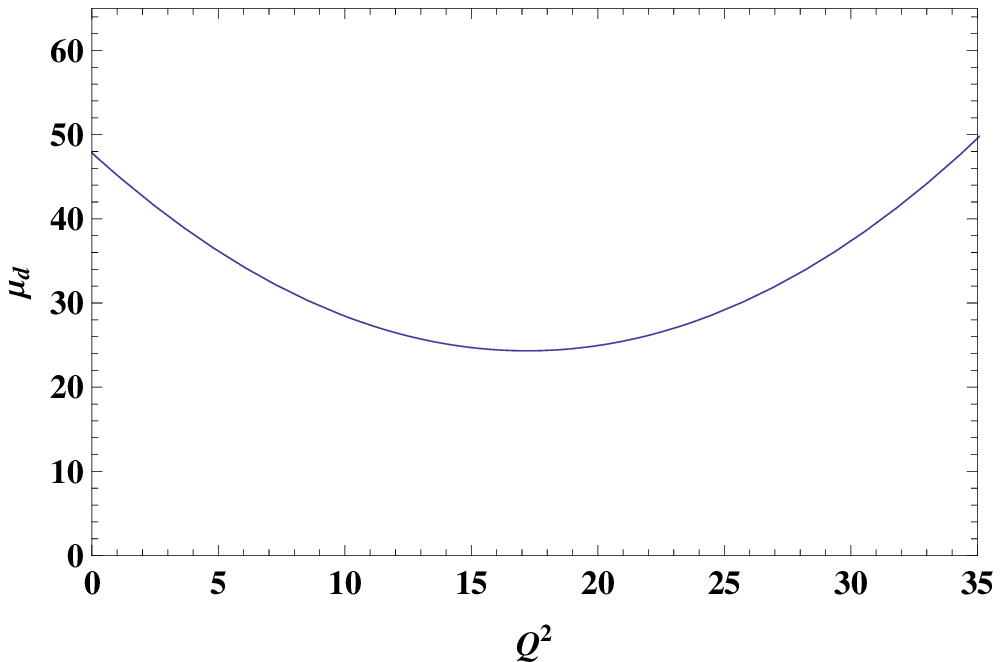,width=6 cm}} \caption{\hspace{-0.05
cm}Chemical potential for down quark, $\mu_d$, verses $Q^2$.
\label{f4}}
\end{minipage}
\end{tabular}
\end{figure}

Therefore we achieve to the temperature, volume and chemical
potentials at different energy scales. Then we are able to fit a
polynomial function to the data points, related to different
values of the statistical parameters  at different energy scales
which are corresponded to the various energy scales of $S_G$ in
[\refcite{Sg}] . We can then obtain energy dependence of the
parameters in the used statistical model (T, V, $\mu_u$, $\mu_d$).
The fitted polynomial function for the concerned parameters as a
function of $Q^2$ is chosen according to the following general
form:
\begin{equation}\label{function2}
g(Q)= p Q^2 + q Q +r \;.
\end{equation}
The  $p$, $q$ and $r$ are the parameters  which are determined by
a fit and can be assigned them the  values as listed in
Table.\ref{ta1}. In Fig. \ref{f1}-\ref{f4} we plot the $V$, $T$,
$\mu_u$ and $\mu_d$ parameters as a function of $Q$, based on the
Eq.(\ref{function2}) .
\begin{table}
\tbl{The numerical values of parameters in Eq.(\ref{function2}).}
{\begin{tabular}{@{}ccccccccccccc@{}} \toprule \hphantom{0000000}
$g(x)$ \hphantom{00000000}\hspace{0.0 cm} V\hphantom{00000000}
T\hphantom{0000000000} $\mu_u$\hphantom{0000000000} $\mu_d$\\
\colrule
\hphantom{0000000000000000}\hspace{0.0 cm}p\hphantom{0000}\hspace{0 cm}-4.27204$\times$$10^{-8}$\hphantom{000}0.0402465\hphantom{000}0.127655\hphantom{00000}0.0794809\\
\hphantom{00000000000000}\hspace{0.0 cm}q\hphantom{000000}\hspace{0 cm}1.45577$\times$$10^{-6}$\hphantom{000}-1.37801\hphantom{000}-4.38716\hphantom{00000}-2.73393\\
\hphantom{000000000000}\hspace{0.0 cm}r\hphantom{000000}\hspace{0 cm}5.77185$\times$$10^{-6}$\hphantom{000}52.9615\hphantom{0000}83.094\hphantom{00000}\hspace{0.0 cm}47.8348\\
 \botrule
\end{tabular} \label{ta1}}
\end{table}
In Fig. \ref{f5}  we plot the $\bar d-\bar u$  at $Q^2$= 54
$GeV^2$ when we use  Eq.(\ref{disfun}) which does not contain
$Q^2$-dependence.   A comparison with this quantity when we assume
a $Q^2$-dependence for related parameters in Eq.(\ref{disfun}),
based on Eq.(\ref{function2}) has also been done. As can be seen
by making a $Q^2$-dependence for the statistical parameters, we
achieve to a little bit better agreement for $\bar d-\bar u$ in
comparison with  the available experimental data \cite{JC,EA}.

\begin{figure}
\begin{center}{\psfig{file=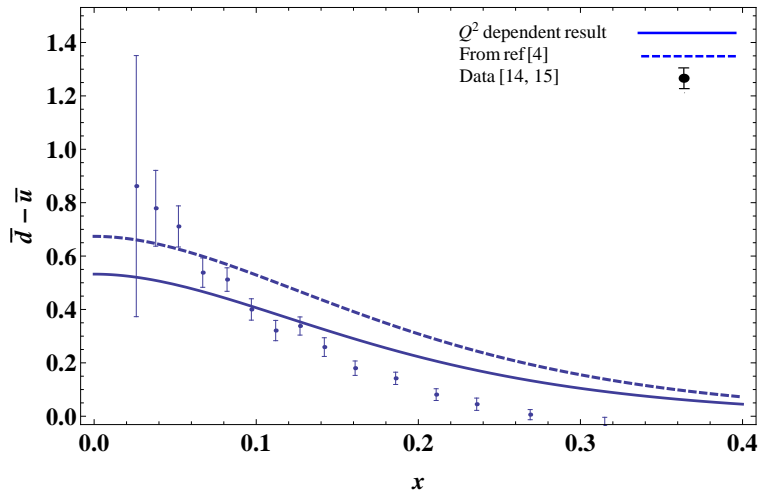,width=10cm}} \caption{ The difference of
${\bar d}$-${\bar u}$, resulted   from Eq.(\ref{disfun}) (dotted
curve). The solid curve is resulted from the the $Q^2$ dependence
of the statistical parameters. Experimental data is quoted from
\protect\cite{JC,EA} \label{f5}}
\end{center}
\end{figure}
\section{Confining potential and the statistical approach}
To obtain the flavor asymmetry in the sea of the nucleon, given by
$\frac{\bar u}{\bar d}$ and ${\bar d}-{\bar u}$, a
phenomenological statistical model presented in \cite{tft,tomio}
is followed with linear confined quarks. In this model the
temperature parameter is given by the $S_G$ violation and the
chemical potentials by the net number of u and d quarks in the
nucleon. To describe the nucleon, two different chemical
potentials are required in which one fixes the net number of u
quarks and the other one the net number of d quarks. The models
studied in \cite{ct,mu} inspire the given  approach in \cite{tft}.
Here the quark energies are not assuming to have continuum levels
as in \cite{ct,mu}. In contrast, a Dirac confining potential
\cite{fhzn} were considered to generate the single-particle
spectrum in which the given quark levels obey the Fermi
statistics. In this case for each quark flavor $\lambda_q$, the
strength of the confining potential $\lambda_q$, is fitted through
the hadron masses where $m_u =m_d = 0$ and it was assumed
$\lambda_u$ = $\lambda_q$ =$\lambda$.

It is not an easy task to develops a quark model with confining
potential to incorporate the  $x$-dependence of quark distribution
functions. By assuming the non-perturbative properties of the
hadron wave function, the recent approach to  incorporate  the
x-dependence of the quark distribution functions is obtained as
suggested in \cite{ai}. In the present statistical quark model,
using  a dynamical input, i.e., the relativistic quark confining
potential, we can achieve to  the $x$-dependence of the
probability functions and the related observable. To extract the
$x$-dependence of the quark amplitudes, the momentum
representation of the quark eigenstates, related  to the Dirac
Hamiltonian is written with respect to the light-cone momenta. The
single-particle nature of the model  makes simple this
transformation.

All the individual quarks of the system including the  valence and
sea quarks in the present statistical quark model are confined by
a central effective interaction. The scalar and vector components
of the related potential with strength $\lambda$, is given by
\cite{fhzn}:

\ba V\left( r \right) = \left( {1 + \beta } \right)\frac{{\lambda
r}}{2}\;. \label{potential}\ea For the u and d quarks, the
strength $\lambda$ of this potential is identical. Considering the
chosen potential in Eq.({\ref{potential}}), an equation similar to
the Schrodinger equation is resulted from the coupled Dirac
equation which can be solved by using the conventional methods of
non-relativistic dynamics. The related Dirac equation which should
be solved is given by \ba \left[ {\vec \alpha .\vec p + \beta m +
V\left( r \right)} \right]{\psi _i(\vec{r})} = {\varepsilon
_i}{\psi _i(\vec{r})}\;.\ea

The $\beta$ and $\vec \alpha$ are the fourth usual $4\times4$
Dirac matrix. They  can be written in terms of the $2 \times 2$
Pauli matrices. Using $\psi _i(\vec{r})$ which is given by
\ba{\psi _i}(\vec{r}) = \left( {\begin{array}{*{20}{c}} 1\\{{{\vec
\sigma .\vec p} \mathord{\left/{\vphantom {{\vec \sigma .\vec p}
{\left( {m + {\varepsilon _i}} \right)}}} \right.
 \kern-\nulldelimiterspace} {\left( {m + {\varepsilon _i}} \right)}}}  \\
\end{array}} \right){\varphi _i}(\vec{r})\;,\ea
the final coupled equations can be transformed to a single second
order differential equation such that: \ba\left[ {{p^2} + \left(
{m + {\varepsilon _i}} \right)\left( {m + \lambda r - {\varepsilon
_i}} \right)} \right]{\varphi _i} = 0\;.\ea Using partial wave
expansion, this equation can be solved for the radial part and in
the case of the s wave $( {\ell = 0})$, where ${j^p} = {\left( {{1
\mathord{\left/
 {\vphantom {1 2}} \right.
 \kern-\nulldelimiterspace} 2}} \right)^ + }$, the radial part of ${\varphi _i}$
 can be written in terms of  the Airy function (Ai): \ba {\varphi _i}\left( r \right) = \sqrt
{\frac{{{k_i}}}{{4\pi }}} \frac{{Ai\left( {{k_i}r + {a_i}}
\right)}}{{{{\left. {r\left[ {\frac{{dAi\left( x \right)}}{{dx}}}
\right]} \right|}_{x = {a_i}}}}}\label{phiai}\;.\ea The  ${a_i}$
is the related {\it i}th root of $Ai\left( x \right)$, ${k_i} =
\sqrt[3]{{\lambda \left( {m + {\varepsilon _i}} \right)}}$, $m$ is
denoting  the current quark mass, finally ${\varepsilon _i}$ are
representing the energy levels and are given by \ba{\varepsilon
_i} = m - \frac{\lambda }{{{k_i}}}{a_i}\;.\label{ep}\ea The
energies for the $u$ and $d$ quarks with $m = 0$ are given by \ba
{\varepsilon _i} = \sqrt \lambda {\left( { - {a_i}}
\right)^{{\raise0.7ex\hbox{$3$} \!\mathord{\left/
 {\vphantom {3 4}}\right.\kern-\nulldelimiterspace}
\!\lower0.7ex\hbox{$4$}}}}\;.\ea
\section{Statistical approach and the  quark model}
Using the relativistic linear confining potential, the statistical
quark model can be  investigated to yield us the nucleon structure
function. Via  this model \cite{pl}, the energy levels for the
quarks, $\varepsilon _i$, can be determined. In this model the
nucleon involves three valence quarks and the sea quarks. Here the
contribution of gluon fields is small and can be neglected
\cite{mac,dey}.

For the quarks in the present statistical quark model, the
Fermi$-$Dirac  distribution is assumed. For a quark system, with
energy levels $\varepsilon _i$ and temperature T, the probability
density for a quark system is given by \ba{\rho _q}\left( \vec{r}
\right) = \sum\limits_i {{g_i}} \psi _i^\dag \left( \vec{r}
\right){\psi _i}\left( \vec{r} \right)\frac{1}{{1 + \exp \left(
{\frac{{{\varepsilon _i} - {\mu _q}}}{T}}
\right)\label{psi}}}\;.\ea The
 $|\psi _i \left( \vec{r}
\right)|^2$ is representing the  density probability for each
state which is  normalized to 1 and the $g_i$ gives the level
degeneracy.

The current quark masses given by $m_u = m_d = 0$ are considered
here which are in fact contains the light quarks. Using a
confining potential model in the Dirac equation, the energies for
the $u$ and $d$ quarks are taken to be equal. With this assumption
the normalization for the proton  is as follows:
\ba\begin{array}{l}
 \int {\left[ {{\rho _q}\left( r \right) - {{\bar \rho }_q}\left( r \right)} \right]} {d^3}r \\
 {\rm{  }} = \sum\limits_i {{g_i}} \left[ {\frac{1}{{1 + \exp \left( {\frac{{{\varepsilon _i} - {\mu _q}}}{T}} \right)}} -
 \frac{1}{{1 + \exp \left( {\frac{{{\varepsilon _i} + {\mu _q}}}{T}} \right)}}} \right] = \left\{ {\begin{array}{*{20}{c}}
   {1{\rm{   }}\; for\; {\rm{  }}q = d}  \\
   {{\rm{2   }}\; for\; {\rm{  }}q = u}  \\
\end{array}} \right. \;.\\
 \end{array}\ea

In order to calculate the nucleon structure function, it is needed
to write  the quark wave function in momentum space, taking the
Fourier transform:

\ba{\Phi _i}\left( {\vec p} \right) = \frac{1}{{{{\left( {2\pi }
\right)}^{{\raise0.7ex\hbox{$3$} \!\mathord{\left/
 {\vphantom {3 2}}\right.\kern-\nulldelimiterspace}
\!\lower0.7ex\hbox{$2$}}}}}}\int {\exp \left( { - i\vec p.\vec r}
\right)} {\psi _i}\left( r \right){d^3}r \label{fouri}\;.\ea
Considering the probability density as
$\varrho=\Phi^{\dagger}_{i}(p)\Phi_{i}(p)$, the quark distribution
in the Bjorken-$x$ space is obtained using two different methods:

In the first method which we call it ``k method" the required
distribution is obtained by the relation \cite{hjprd,MM}:
\begin{eqnarray}
q(x)=2\pi M_{t}\int^{\infty}_{k_{min}}\varrho p
\textmd{d}p\;,\label{parton}\end{eqnarray} where
\begin{eqnarray}k_{min}=\frac{xM_{t}+\varepsilon_{i}}{2}-\frac{m^{2}}{2(xM_{t}+\varepsilon_{i})}\nonumber \;.
\end{eqnarray}
In this equation $M_{t}$ and $m$ are denoting the proton and $u,d$
quark masses respectively. The energy level of ground state is
given by $\varepsilon_{i}$.

In the second method which we call it ``null-plane method" , to
render the parton distribution in the Bjorken $x$-space, the quark
distribution function is obtained by the relation \cite{Ma 08}:
\begin{equation}
q(x)=\int q(x,\vec{p}_{\perp})\textmd{d}^{2}p_{\perp}\;,\label{fx}
\end{equation}
where $q(x,\vec{p}_{\perp})$ satisfies the following relation:
\begin{equation}
\int\varrho\delta(p^{0}-\sqrt{(p^{3})^{2}+(\vec{p}_{\perp})^{2}+m^{2})}\textmd{d}p^{0}\textmd{d}p^{3}\textmd{d}^{2}p_{\perp}=
\int q(x,\vec{p}_{\perp})\textmd{d}^{2}p_{\perp}\textmd{d}x\;.
\label{integer}
\end{equation}
Note that when doing the integration in Eq.(\ref{integer}), the
on-shell condition $p^{0}=\sqrt{\vec{p}^{2}+m^{2}}$ is needed,
where $p^{0}$, $\vec{p}=(p^{1},p^{2},p^{3})$ and  $m$ are the
energy, 3-momentum and mass of the quark respectively. We should
emphasize that the ``null-plane method" which was also used in
Refs.[\refcite{tomio,tft}] was the main motivation to lead us to
follow the calculations based on the first method.

In brief as was dealt  with in Sect. 3, using the null plane
variables, we will have \ba\begin{array}{l}
{p^ + } = x{P^ + }{\rm{,}}\;\;{P^ + } = {M_t}{\rm{   }}\left( {t = p,n} \right) \\
{p_z} = {p^ + } - {\varepsilon _i} = {M_t}\left( {x - \frac{{{\varepsilon _i}}}{{{M_t}}}} \right) \\
\end{array}\;,\ea
where $x$ is the momentum fraction of the nucleon carried by the
quark, $M_t$ is the nucleon thermal mass at a given temperature T.
Therefore we redefine the wave function in  Eq.(\ref{fouri}) as

\ba{\Phi _i}\left( {\vec p} \right) = {\Phi _i}\left( {x,{p_ \bot
}} \right)\label{phiii}\;.\ea By integrating Eq.(\ref{psi}) over
${p_ \bot }$ where $\psi(\vec r)$ is replaced by ${\Phi _i}\left(
{x,{p_ \bot }} \right)$  we obtain the quark structure function
for each flavor q as in the following:

\ba{q_T}\left( x \right) = \sum\limits_i {\int {{d^2}{p_ \bot }} }
\frac{{\Phi _i^\dag {\Phi _i}}}{{1 + \exp \left(
{\frac{{{\varepsilon _i}
 - {\mu _q}}}{T}} \right)}}\label{phi}\label{qt}\;.\ea

In fact for ${\Phi _i}$ in Eq.(\ref{phiii}) we have:

\ba{\Phi _i} = {\Phi _i}\left( {{M_t}\left( {x -
\frac{{{\varepsilon _i}}}{{{M_t}}}} \right),{p_ \bot }}
\right)\;.\ea

In Eq.(\ref{qt}), ${q_T}\left( x \right)$ describes the
probability that a quark with flavor $q$ has a fraction $x$ of the
total momentum of the nucleon, assuming a temperature T . For the
corresponding anti-quark distribution, $\bar q_T\left( x \right)$,
we have to replace $\mu_q$ by $\mu_{\bar{q}}$ in Eq.(\ref{qt}).

\section{An effective approach to the linear confining potential}
In a standard procedure, we should take massless quarks and equal
potentia strength for different flavor quarks and then to do a sum
over different energy levels as in  Eq.(\ref{qt}) to achieve to
the quark distributions as a function of Bjorken-$x$ variable.
What we do in this section is  different with respect to what was
assumed in previous section. Here  we assign different masses to
$u$ and $d$ quarks so as: $m_u$=187 $MeV$, $m_d$=196 $MeV$ as
current quark masses. The strength of linear potential which is
denoted by $\lambda$  in Eq.(\ref{potential}) is assumed identical
for $u$ and $d$ quarks in which we take into account $\lambda$=239
$MeV$ in correspond to what was quoted in \cite{tft}. We assume
just one energy level effectively and due to the  different  quark
mass, the amount of energy level is not identical  for different
quark flavors so as we take ${\varepsilon _u}$=290 $MeV$ and
${\varepsilon _d}$=225 $MeV$ for $u$ and $d$ quarks respectively.
This will lead us to the different values for $k_i$ in
Eq.(\ref{ep}). In our calculations the second root of Airy
function is used, $a_2$=-4.08798, which is more appropriate for
taken effective approach in this section.

As was pointed out in previous section to convert  the quark wave
function from momentum space to Bjorken-$x$ space, we employ the
second method which was called ``k method''. The integral in
Eq.(\ref{fouri}) is a triple integral. The integral over azimuthal
variable can be done strictly due to cylindrical symmetry. The
integration over polar variable from $\theta$=0 to $\theta$=$\pi$
can be easily calculated. The final integration over radial
variable, $r$, can also be done analytically. By replacing the
final result for $\Phi_i(p)$ in Eq.(\ref{parton}), we can obtain
the quark distributions in terms of the $x$-variable. If one
wishes to do the integration in Eq.(\ref{parton}) analytically, it
will be very hard so we prefer to do this integration numerically.
For this propose, we need to make a data table which contains two
columns. First column includes the different values of $p$
variable from $p$=-a to $p$=a where a is  a large number. The
chosen interval for $p$ is due to this fact that the $\Phi_i (p)$
is an even function with respect to $p$. The second column is
containing the numerical values of $p\varrho$
($\varrho=\Phi^{\dagger}_{i}(p)\Phi_{i}(p)$), considering the
different values of $p$-values in the first column. By considering
the shape of the  $p\varrho$ combination with respect $p$  which
can be obtained by a list plot of extracted data, a proper fitting
function for the concerned quantity can be conjectured. We take
the following function for the fit: \ba
f(p)=a\;p\;e^{-b\;p^2}+c\;p^3\;e^{-d\;p^4}+h\;p\;e^{-g\;p^6}\label{fit}\;,\ea
where $a, b, c, d, h$ and $g$ are the fitting parameters.  By
substituting the fitted parameters for $u$ and $d$ quark
respectively in Eq.(\ref{fit}) and then back the result into the
Eq.(\ref{parton}), the related quark distribution in $x$-space
will be obtained. The numerical value for $M_t$  in this equation
is taken to equal to $M_t$=985 $MeV$. In this stage of
calculations, we can normalize the bare quark distributions for
$u$ and $d$ quarks to 2 and 1 as are required.

The next step is to impose  this result on the employed
statistical effect by multiplying it with the related factor in
Eq.(\ref{qt}), that is:

\ba  \frac{1}{{1 + \exp \left( {\frac{{{\varepsilon _i}
 - {\mu _q}}}{T}} \right)}}\label{phi}\label{qqt}\;.\ea
Then by putting the final results in the sum rules, given by
Eqs.(\ref{sumrule1},\ref{sumrule2}) and Eq.(\ref{gottfried}), we
can obtain the unknown parameters $\mu_u$, $\mu_d$ and $T$ by
solving simultaneously the related set of equations. The numerical
values for these parameters which we obtain, are : $\mu_u$=908.42
$MeV$, $\mu_d$=720.094 $MeV$ and $T$=219.56. Since the
calculations are done effectively, the numerical values for
temperature and chemical potentials would be different with
respect to what are existed in  the usual statistical model. The
essential point which we should take into account is that the
numerical behavior of  these quantities is like the one which  is
expecting from   the usual statistical model.

By substituting these numerical values in Eq.(\ref{qt}) and
choosing the proper sign and values for the chemical potential, we
can achieve to $\bar u$ and $\bar d$ distributions. The results
for $\frac{\bar d}{\bar u}$ and $\bar d- \bar u$ at $Q^2$=54
$GeV^2$ are depicted in Fig.6 and Fig.7. The comparison with the
available experimental data \cite{JC,EA}  and the result from
Ref.[\refcite{tomio}] has also been done there.

\begin{figure}
\begin{center}{\psfig{file=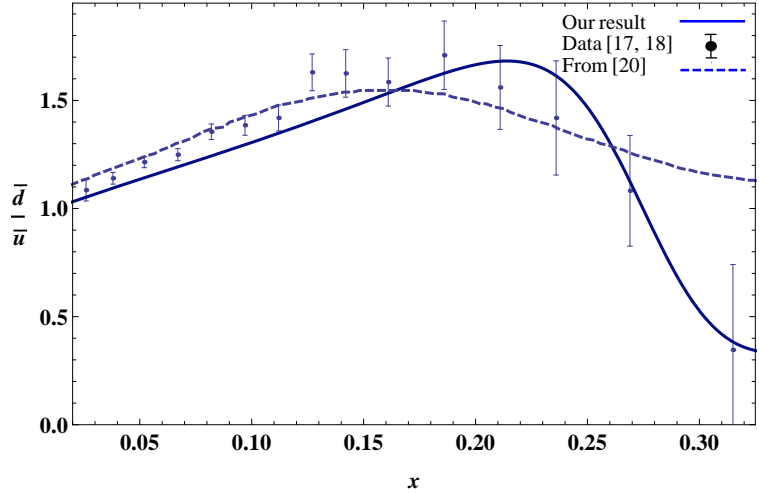,width=10cm}} \caption{ The
ratio of $\frac{\bar d}{\bar u}$, resulted   from
Eq.(\ref{qt})(solid curve) in an effective approach. Experimental
data is quoted from \protect\cite{JC,EA}. Comparison with the
result from [\protect\refcite{tomio}] (dash-dotted curve) has also
been done. \label{f6}}\end{center}
\end{figure}

\begin{figure}
\begin{center}{\psfig{file=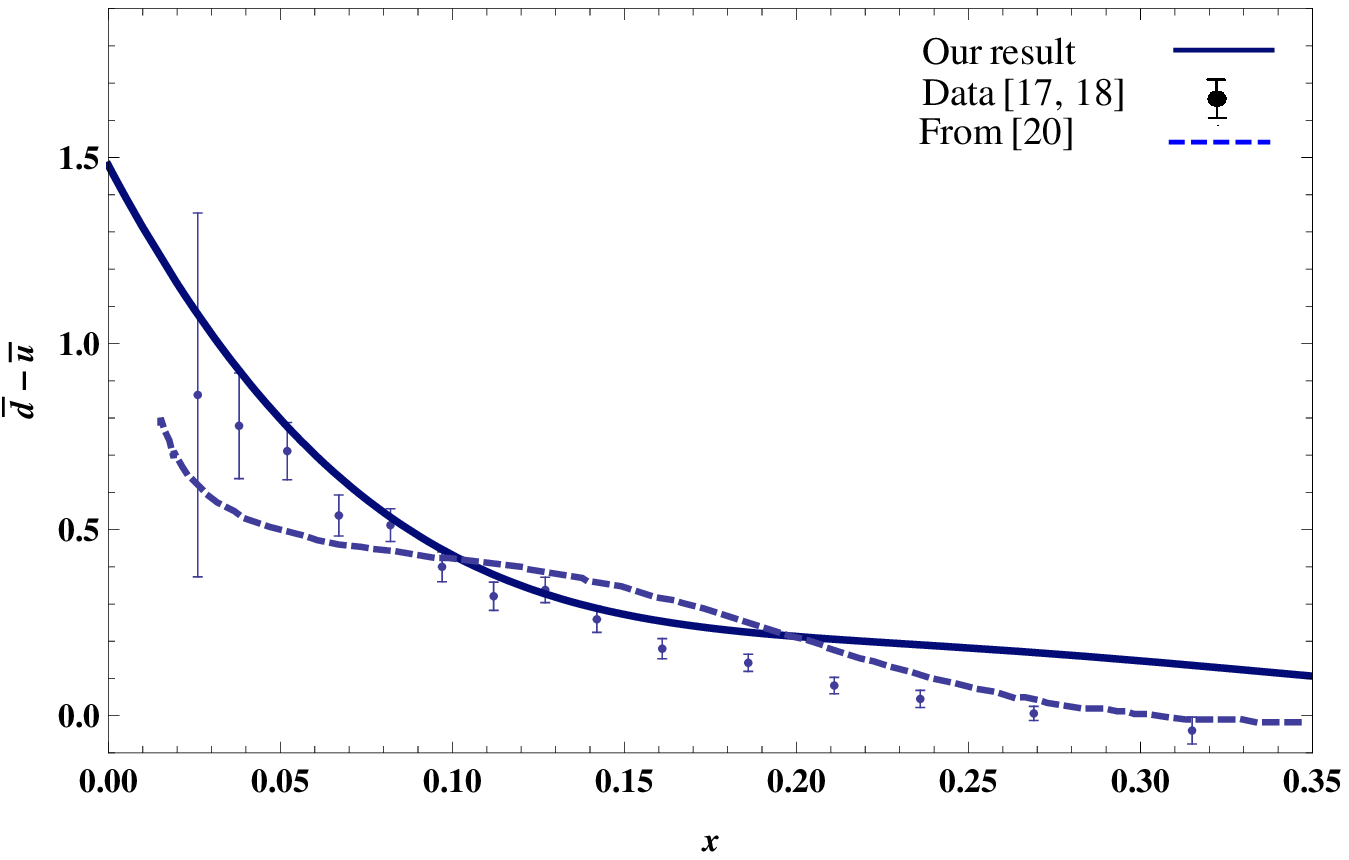,width=10cm}} \caption{ The
difference of ${\bar d}$-${\bar u}$, resulted   from
Eq.(\ref{qt})(solid curve) in an effective approach. Experimental
data is quoted from Refs.[\protect\refcite{JC,EA}]. Comparison
with the result from [\protect\refcite{tomio}] (dash-dotted curve)
 has also been done. \label{f7}}\end{center}
\end{figure}

As we mentioned at the beginning  of this section we use
effectively the confining linear potential to indicate that the
model is working well. In fact we intend to show that this model
has inherently this ability to give us an acceptable result for
quark distributions. We get  these results by choosing proper
numerical values for quantities like $m_u$, $m_d$, ... as we refer
them at the first part of this section. Using these values as we
explained before, we extract the $\frac{\bar d}{\bar u}$ and $\bar
d- \bar u$ in a good agreement with the available experimental
data and even   in a  better agreement  (more and less) with the
results from Ref.[\refcite{tomio}]. Once again we emphasize that
our approach is an effective approach and is not in complete
correspondence  to what described in previous section.

As can be seen in \cite{tomio}, just using the linear confining
potential of Sect.6 with the assumptions which existed there, will
not give us the  proper results for the $\frac{\bar d}{\bar u}$
and $\bar d- \bar u$. In fact to achieve to the proper results,
following the strategy of Sect.6, we should add some extra effects
to the calculations, such as mass shift, instanton effect and
quark substructure like gluonic and pionic effects as were used in
\cite{tomio}. Considering these extra effects  will yield   the
proper results for the concerned quantities. But we show in this
section that we can obtain the proper results without using these
extra effects, just by following  an effective approach which we
describe here, using chosen numerical values for the required
quantities.

\section{Conclusions}
Here we had a review on the statistical approach based on the Ma
{\it et al.}'s article\cite{ma2009} to obtain the parton densities
inside the nucleon. In further step we did dependence the related
statistical parameters on $Q^2$ by using the different values of
$S_G$ at different values of $Q^2$ in correspond to neural method
in Ref.[\refcite{Sg}]. In this case the result for $\bar d$-$\bar
u$ at $Q^2$= 54 $GeV^2$ would be in better  agreement with the
available experimental data \cite{JC,EA} as can be seen in Fig.5.

So far the partons were assumed as free particles. In real case
the interaction between partons should be considered which has
been done in Sect.6 by taking into account a linear confining
potential between the quarks. Employing this interaction  would
improve the results for the extracted parton densities with
respect to the case where the partons were taken as free
particles. For this propose, at first the wave function of quarks
were  obtained in spatial coordinates, using the  solution of  the
Dirac equation under the linear confining potential. In
continuation, the quark wave function should be converted to
momentum space which could be done by a Fourier transformation.
Later on this wave function would be appeared in Bjorken $x$-space
by two methods which were called ``null-plane method" and ``k
method" as were described in Sect.6. We used in this article the
``k method" method to obtain the quark distributions in $x$-space.

According to the descriptions of Sect.6, the real quark
distribution would be obtained by doing a summation over different
energy levels of quarks in confining potential while massless
quarks were assumed. In the effective approach which we took into
account  in Sect.7, we assumed massive quarks with one  energy
level which were different for $u$ and $d$ quarks. What we got in
this approach were acceptable and in a good agreement with $\bar
d$-$\bar u$ and the ratio $\frac{\bar d}{\bar u}$ at $Q^2$= 54
$GeV^2$ in comparison with the available experimental data and the
obtained result from \cite{tomio}. This showed that the linear
confining potential contains alone and inherently this ability to
produce proper results for parton densities without resorting to
some extra effects like mass shift and etc. as were done in
Ref.[\refcite{tomio}].

Considering other confining potential like M.I.T bag model or
squared radial potential would also  be interesting to yield us
the parton densities in $x$-space. We hope to do in this
connection some reports in our further research job. Extending the
calculations to the polarized case would  be as well an attractive
subject to follow it  as our new scientific task.
\appendix
\section{}

We indicate here the details of numerical calculations, based on
Eq.(\ref{parton}) which lead us to parton densities, using linear
confining potential.

Employing the  numerical values for quark  binding energies and
the other numerical values for the required quantities as are
indicated in the article, the result of integration  in
Eq.(\ref{fouri}) with respect to $p$ variable for the $u$ quark is
as follows:

\ba &&\Phi_u(p)=\frac{1}{p^3}\;0.00137543\;i\;e^{9.95192\; i\;
p-4.80931\; i\;
p^3}(-21.7656\; i \sqrt[3]{-i p^3}\; p-21.7656\; (-i \;p^3)^{2/3}\nonumber\\
&&+16.0736\; i\; \sqrt[3]{-i p^3}\; p\; \Gamma
(\frac{2}{3},-4.80931 i\; p^3)+8.12471 (-i p^3)^{2/3}\; \Gamma
(\frac{1}{3},-4.80931\; i p^3)+4 \sqrt{3} \pi p^2) \nonumber\\ &&
-\frac{1}{p^3}\;0.00137543\; i\; e^{4.80931\; i\; p^3-9.95192\;
i\; p}\; (21.7656\; i \sqrt[3]{i p^3} \;p-21.7656 \;(i\;
p^3)^{2/3}\nonumber\\
&&-16.0736 \;i \;\sqrt[3]{i p^3}\; p\Gamma (\frac{2}{3},4.80931\;
i\; p^3)+8.12471\; (i\; p^3)^{2/3} \Gamma(\frac{1}{3},4.80931\;
i\; p^3)+4\; \sqrt{3}\; \pi\; p^2)\;.\nonumber\\ \ea For the $d$
quark we will have:
\begin{table}
\tbl{The amounts of
$p\varrho_i$=$p\Phi^{\dagger}_{i}(p)\Phi_{i}(p)$ versus different
values of $p$.} {\begin{tabular}{@{}ccccc@{}} \toprule
$p$\hphantom{00000} & \hphantom{00000}
$p\Phi^{\dagger}_{u}(p)\Phi_{u}(p)$
& \hphantom{00000} $p\Phi^{\dagger}_{d}(p)\Phi_{d}(p)$\\
\colrule
\hphantom{00}-0.2\hphantom{00000}\hphantom{00}  &\hphantom{00000}-0.00799423&\hphantom{00000}-0.00429547\\
\hphantom{00}-0.19\hphantom{00000}\hphantom{00} &\hphantom{00000}-0.0104922 &\hphantom{00000}-0.00645628\\
\hphantom{00}-0.18\hphantom{00000}\hphantom{00} &\hphantom{00000}-0.0131948&\hphantom{00000}-0.00899122\\
\hphantom{00}-0.17\hphantom{00000}\hphantom{00} &\hphantom{00000}-0.0160138&\hphantom{00000}-0.0118213\\
\hphantom{00}-0.16\hphantom{00000}\hphantom{00} &\hphantom{00000}-0.0188475&\hphantom{00000}-0.0148434\\
\hphantom{00}-0.15\hphantom{00000}\hphantom{00} &\hphantom{00000}-0.0215853&\hphantom{00000}-0.0.017934\\
\hphantom{00}-0.14\hphantom{00000}\hphantom{00} & \hphantom{00000}-0.0241117&\hphantom{00000}-0.0209543\\
\hphantom{00}-0.13\hphantom{00000}\hphantom{00} &\hphantom{00000}-0.0263111 &\hphantom{00000}-0.0237567\\
\hphantom{00}-0.12\hphantom{00000}\hphantom{00} &\hphantom{00000}-0.0280731&\hphantom{00000}-0.0261918\\
\hphantom{00}-0.11\hphantom{00000}\hphantom{00} &\hphantom{00000}-0.0292978&\hphantom{00000}-0.0.0281166\\
\hphantom{00}-0.10\hphantom{00000}\hphantom{00} &\hphantom{00000}-0.0299005&\hphantom{00000}-0.0294021\\
\hphantom{00}-0.09\hphantom{00000}\hphantom{00} &\hphantom{00000}-0.0298161&\hphantom{00000}-0.0.029941\\
\hphantom{00}-0.08\hphantom{00000}\hphantom{00} & \hphantom{00000}-0.0290029&\hphantom{00000}-0.029654\\
\hphantom{00}-0.07\hphantom{00000}\hphantom{00} &\hphantom{00000}-0.0274451 &\hphantom{00000}-0.0284955\\
\hphantom{00}-0.06\hphantom{00000}\hphantom{00} &\hphantom{00000}-0.0251548&\hphantom{00000}-0.0264573\\
\hphantom{00}-0.05\hphantom{00000}\hphantom{00} &\hphantom{00000}-0.0221718&\hphantom{00000}-0.0235702\\
\hphantom{00}-0.04\hphantom{00000}\hphantom{00} &\hphantom{00000}-0.018563&\hphantom{00000}-0.0199035\\
\hphantom{00}-0.03\hphantom{00000}\hphantom{00} &\hphantom{00000}-0.0144202&\hphantom{00000}-0.0155633\\
\hphantom{00}-0.02\hphantom{00000}\hphantom{00} &\hphantom{00000}-0.00985654&\hphantom{00000}-0.0106873\\
\hphantom{00}-0.01\hphantom{00000}\hphantom{00} &\hphantom{00000}-0.00500241&\hphantom{00000}-0.00543901\\
\hphantom{00}0.00\hphantom{00000}\hphantom{00}  &\hphantom{00000}0.00000004 &\hphantom{00000}0.000000003\\
\hphantom{00}0.01\hphantom{00000}\hphantom{00}  &\hphantom{00000}0.00500241 &\hphantom{00000}0.00543901\\
\hphantom{00}0.02\hphantom{00000}\hphantom{00}  &\hphantom{00000}0.00985654 &\hphantom{00000}0.0106873\\
\hphantom{00}0.03\hphantom{00000}\hphantom{00}  &\hphantom{00000}0.0144202&\hphantom{00000}0.0155633\\
\hphantom{00}0.04\hphantom{00000}\hphantom{00}  &\hphantom{00000}0.018563&\hphantom{00000}0.0199035\\
\hphantom{00}0.05\hphantom{00000}\hphantom{00}  &\hphantom{00000}0.0221718&\hphantom{00000}0.0235702\\
\hphantom{00}0.06\hphantom{00000}\hphantom{00}  &\hphantom{00000}0.0251548 &\hphantom{00000}0.0264573\\
\hphantom{00}0.07\hphantom{00000}\hphantom{00}  &\hphantom{00000}0.0274451&\hphantom{00000}0.0284955\\
\hphantom{00}0.08\hphantom{00000}\hphantom{00}  &\hphantom{00000}0.0290029&\hphantom{00000}0.029654\\
\hphantom{00}0.09\hphantom{00000}\hphantom{00}  &\hphantom{00000}0.0298161&\hphantom{00000}0.029941\\
\hphantom{00}0.10\hphantom{00000}\hphantom{00}  &\hphantom{00000}0.0299005&\hphantom{00000}0.0294021\\
\hphantom{00}0.11\hphantom{00000}\hphantom{00}  &\hphantom{00000}0.0292978&\hphantom{00000}0.0281166\\
\hphantom{00}0.12\hphantom{00000}\hphantom{00}  &\hphantom{00000}0.0280731 &\hphantom{00000}0.0261918\\
\hphantom{00}0.13\hphantom{00000}\hphantom{00}  &\hphantom{00000}0.0263111&\hphantom{00000}0.0237567\\
\hphantom{00}0.14\hphantom{00000}\hphantom{00}  &\hphantom{00000}0.0241117&\hphantom{00000}0.0209543\\
\hphantom{00}0.15\hphantom{00000}\hphantom{00}  &\hphantom{00000}0.0215853&\hphantom{00000}0.017934\\
\hphantom{00}0.16\hphantom{00000}\hphantom{00}  &\hphantom{00000}0.0188475&\hphantom{00000}0.0148434\\
\hphantom{00}0.17\hphantom{00000}\hphantom{00}  &\hphantom{00000}0.0160138&\hphantom{00000}0.0118213\\
\hphantom{00}0.18\hphantom{00000}\hphantom{00}  &\hphantom{00000}0.0131948&\hphantom{00000}0.00899122\\
\hphantom{00}0.19\hphantom{00000}\hphantom{00}  &\hphantom{00000}0.0104922&\hphantom{00000}0.00645628\\
\hphantom{00}0.20\hphantom{00000}\hphantom{00}  &\hphantom{00000}0.00799423&\hphantom{00000}0.00429547\\
\botrule
\end{tabular} \label{ta1}}
\end{table}
\ba &&\Phi_d(p)=\frac{1}{p^3}\;0.00131846\; i\; e^{10.8304 i
p-6.19867\; i\; p^3} (-21.7656\; i \;\sqrt[3]{-i p^3}\; p-21.7656
\;(-i p^3)^{2/3}\nonumber\\&&+16.0736\; i\; \sqrt[3]{-i p^3}\; p\;
\Gamma (\frac{2}{3},-6.19867\; i\; p^3)+8.12471 \; (-i\;
p^3)^{2/3} \Gamma (\frac{1}{3},-6.19867\; i \;p^3)\nonumber\\
&& +4 \sqrt{3}\; \pi\; p^2)-\frac{1}{p^3}\;0.00131846\; i\;
e^{6.19867\; i\; p^3-10.8304 \;i\; p} (21.7656\; i\; \sqrt[3]{i
p^3}\; p-21.7656 \;(i\; p^3)^{2/3}\nonumber\\ &&-16.0736\; i\;
\sqrt[3]{i p^3}\; p \Gamma (\frac{2}{3},6.19867\; i\;
p^3)+8.12471\; (i\; p^3)^{2/3}\; \Gamma (\frac{1}{3},6.19867\; i\;
p^3)+4\; \sqrt{3}\; \pi\; p^2)\;.\nonumber\\
 \ea
In order to calculate the integration in Eq.(\ref{parton}), we
resort to a numerical method. For this propose we need first to
make a table of data, inclining two columns: one the $p$ variable
in a symmetrical interval, say for instance from -2 to +2 and the
other one is the integrated quantity in Eq.(\ref{parton}). In the
obtained table, the chosen step is 0.1. In practice, the amount of
the step can be chosen less than the assumed step. To avoid of
presenting two separate tables, we merge in related table  the
data for $u$ and $d$ quarks.

Considering the shape of the data, the best function to fit the
data for $u$ quark, is as following: \ba
f_{u}(p)=a_{u}\;p\;e^{-b_{u}\;p^2}+c_{u}\;p^3\;e^{-d_{u}\;p^4}+h_{u}\;p\;e^{-g_{u}\;p^6}\label{fit1}\;,\ea
Fisting the above equation to the related data will lead us to the
following numerical values for the unknown parameters of the
equation: \ba &&a_{u}\rightarrow
0.502886,\;\;\;\;\;b_{u}\rightarrow
46.3325,\;\;\;\;\; c_{u}\rightarrow -1.77947,\nonumber\\
&&d_{u}\rightarrow 379.021,\;\;\;\;h_{u}\rightarrow
-2.88708,\;\;\;\; g_{u}\rightarrow 122054.\ea

For the $d$ quark, the fitted equation is like the $u$ one, so as:

\ba
f_{d}(p)=a_{d}\;p\;e^{-b_{d}\;p^2}+c_{d}\;p^3\;e^{-d_{d}\;p^4}+h_{d}\;p\;e^{-g_{d}\;p^6}\label{fit2}\;,\ea
The numerical results for the unknown parameters are as following:
\ba &&a_{d}\rightarrow 0.547355,\;\;\;\;\;b_{d}\rightarrow
55.0649,\;\;\;\;\; c_{d}\rightarrow -2.22766,\nonumber\\
&&d_{d}\rightarrow 516.719,\;\;\;\;h_{d}\rightarrow
-4.47365,\;\;\;\; g_{d}\rightarrow 148614.\ea\\

The desired form of  Eq.(\ref{parton}) for $u$ quark is appeared
as:
\begin{eqnarray}
u(x)=N_u\; 2\pi M_{t}\int^{\infty}_{k_{min_u}}\varrho_u p
\textmd{d}p\;,\label{parton1}\end{eqnarray} where
\begin{eqnarray}k_{min_u}=\frac{xM_{t}+\varepsilon_{u}}{2}-\frac{m_u^{2}}{2(xM_{t}+\varepsilon_{u})}\nonumber
\; \texttt{and}\;\; \varrho_u =\Phi^{\dagger}_{u}(p)\Phi_{u}(p)
\;.
\end{eqnarray}
In this equation, $N_u$ is a normalization constant to fulfil the
required sum rule for the $u$ quark. Substituting the final form
of Eq.(\ref{fit1}) which is representing $\varrho_u p$,  in
Eq.(\ref{parton1}) will yield us the $u$ quark distribution in the
Bjorken $x$ space:

\ba && u(x)= 8515.86 (-1.779 (0.0006595-\frac{1}{(0.985 x+0.29)^4}
e^{-\frac{20.991 (x+0.104569)^4 (x+0.484264)^4}{(0.985
x+0.29)^4}}\nonumber\\ && (0.000620901 e^{\frac{20.991
(x+0.104569)^4 (x+0.484264)^4}{(0.985 x+0.29)^4}}-0.000620901)
(x+0.294374) (x+0.294458)\nonumber\\ && (x^2+0.588832
x+0.0866809))+(0.502886 (e^{-\frac{10.9036 (x+0.104569)^2
(x+0.484264)^2}{(0.985 x+0.29)^2}}\nonumber\\ && ((0.0107915
x+0.00635442) x+0.000935422)-1.787959984515764\times
10^{-18}\nonumber\\ && (x+0.25)^2))/(1.
x+0.294416)^2-0.00000394234\; e^{-\frac{1590.77
(x+0.104569)^6 (x+0.484264)^6}{(0.985 x+0.29)^6}}\nonumber\\
&&-6.113633443709904\times10^{-22}) \ea

Similarly for $d$ quark, the required substitution should be done
in following integration:
\begin{eqnarray}
d(x)=N_d\; 2\pi M_{t}\int^{\infty}_{k_{min_d}}\varrho_d p
\textmd{d}p\;,\label{parton2}\end{eqnarray} where
\begin{eqnarray}k_{min_d}=\frac{xM_{t}+\varepsilon_{d}}{2}-\frac{m_d^{2}}{2(xM_{t}+\varepsilon_{d})}\nonumber
\; \texttt{and}\;\; \varrho_d =\Phi^{\dagger}_{d}(p)\Phi_{d}(p)
\;.
\end{eqnarray}\\\\\\\\
The result of integration in Eq.(\ref{parton2}) for $d$ quark in
$x$ space, would be as in following. As before the $N_d$ is the
required normalization constant. \ba && d(x)= 2573.23 (-2.2276
(0.00048382-\frac{1}{(0.985 x+0.225)^4} e^{-\frac{28.617
(x+0.0294416)^4 (x+0.427411)^4}{(0.985 x+0.225)^4}} \nonumber\\
&&(0.000455439 e^{\frac{28.617 (x+0.0294416)^4
(x+0.427411)^4}{(0.985 x+0.225)^4}}-0.000455439) (x+0.228385)
(x+0.228468) \nonumber\\
&&(x^2+0.456853 x+0.0521786))+(0.547355
(3.575919969031528\times10^{-18} \nonumber\\
&&(x+0.125) (x+0.375)+e^{-\frac{12.9586 (x+0.0294416)^2
(x+0.427411)^2}{(0.985 x+0.225)^2}}\nonumber\\
&& ((0.0090802 x+0.00414831) x+0.000473792)))/(1.
x+0.228426)^2-0.000005017\nonumber\\
&& e^{-\frac{1936.93 (x+0.0294416)^6 (x+0.427411)^6}{(0.985
x+0.225)^6}}-1.2315314835401511\times10^{-20})\ea

By achieving to $u$ and $d$ type quark distributions, all the
desired combination of parton densities in the statistical model
can be obtained.

\end{document}